\documentclass{optica-article}

\journal{opticajournal} 

\articletype{Research Article}
\pdfoutput=1
\usepackage{filecontents}
\usepackage{lineno}
\usepackage[
  toc,
  acronym,
  style = long
]{glossaries}
\usepackage{xr}
\newacronym{lfm}{LFM}{light field microscopy}
\glsadd{lfm}
\newacronym{lfmic}{LFMic}{light field microscope}
\glsadd{lfmic}
\newacronym{lf}{LF}{light field}
\glsadd{lf}
\newacronym{flfm}{FLFM}{Fourier light field microscopy}
\glsadd{flfm}
\newacronym{flfmic}{FLFMic}{Fourier light field microscope}
\glsadd{flfmic}
\newacronym{xlfm}{XLFM}{extended field-of-view light field microscope}
\glsadd{xlfm}
\newacronym{plfm}{PFLFM}{polarization light field microscopy}
\glsadd{plfm}
\newacronym{plfmic}{PFLFMic}{polarization light field microscope}
\glsadd{plfmic}
\newacronym{mla}{MLA}{micro lens array}
\glsadd{mla}
\newacronym{ml}{ML}{micro lens}
\glsadd{ml}
\newacronym{psf}{PSF}{point spread function}
\glsadd{psf}
\newacronym{slm}{SLM}{spatial light modulator}
\glsadd{slm}
\newacronym{dl}{DL}{deep learning}
\glsadd{dl}
\newacronym{rl}{RL}{Richardson-Lucy}
\glsadd{rl}
\newacronym{bnns}{BNNs}{Bayesian neural networks}
\glsadd{bnns}
\newacronym{ood}{OOD}{out-of-distribution}
\glsadd{ood}
\newacronym{oodd}{OODD}{out-of-distribution detection}
\glsadd{oodd}
\newacronym{dsd}{DSD}{domain shift detection}
\glsadd{dsd}
\newacronym{nf}{NF}{normalizing flow}
\glsadd{nf}
\newacronym{nfs}{NFs}{Normalizing Flows}
\glsadd{nfs}
\newacronym{inn}{INN}{Invertible Neural Network}
\glsadd{inn}
\newacronym{wf}{WF}{Wavelet-Flow}
\glsadd{wf}
\newacronym{cwf}{CWF}{Conditional Wavelet-Flow}
\glsadd{cwf}
\newacronym{cwfa}{CWFA}{Conditional Wavelet-Flow Architecture}
\glsadd{cwfa}
\newacronym{cnf}{CNF}{Conditional Normalizing-Flow}
\glsadd{cnf}
\newacronym{cnn}{CNN}{Convolutional Neural Network}
\glsadd{cnn}
\newacronym{nll}{NLL}{negative-log-likelihood}
\glsadd{nll}
\newacronym{ll}{LL}{log-likelihood}
\glsadd{ll}
\newacronym{sd}{SD}{sparse decomposition}
\glsadd{sd}
\newacronym{wfr}{WF}{wave-front}
\glsadd{wfr}
\newacronym{mtf}{MTF}{modulation transfer function}
\glsadd{mtf}
\newacronym{cat}{CAT}{conditional affine transform}
\glsadd{cat}

\newacronym{gpu}{GPU}{graphics processing unit}
\glsadd{gpu}
\newacronym{gt}{GT}{ground truth}
\glsadd{gt}
\newacronym{roi}{ROI}{region of interest}
\glsadd{roi}
\newacronym{3d}{3D}{3-dimensional}
\glsadd{3d}
\newacronym{fwhm}{FWHM}{full with at half maximum}
\glsadd{fwhm}
\newacronym{sdlfm}{SDLFM}{sparse decomposition light field microscopy}
\glsadd{sdlfm}
\newacronym{mape}{MAPE}{mean absolute percentage error}
\glsadd{mape}
\newacronym{psnr}{PSNR}{peak signal-to-ratio}
\glsadd{psnr}
\newacronym{pcc}{PCC}{Pearson correlation coefficient}
\glsadd{pcc}

\newcommand{\eg}{\emph{eg.}}

\begin{document}

\title{Fast light-field 3D microscopy with out-of-distribution detection and adaptation through Conditional Normalizing Flows}

\author{Josué Page Vizcaíno,\authormark{1,2,*} Panagiotis Symvoulidis,\authormark{3}, Zeguan Wang\authormark{3}, Jonas Jelten\authormark{1,2}, Paolo Favaro\authormark{4}, Edward S. Boyden\authormark{3}, Tobias Lasser\authormark{1,2}}

\address{\authormark{1}Computational Imaging and Inverse Problems, Department of Informatics, School of Computation, Information and Technology, Technical University of Munich, Germany\\
\authormark{2}Munich Institute of Biomedical Engineering, Technical University of Munich, Germany\\
\authormark{3}Synthetic Neurobiology Group, Massachusetts Institute of Technology, USA\\
\authormark{4}Computer Vision Group, University of Bern, Switzerland}

\email{\authormark{*}pv.josue@gmail.com} 


\begin{abstract*} 
Real-time 3D fluorescence microscopy is crucial for the spatiotemporal analysis of live organisms, such as neural activity monitoring. The eXtended field-of-view light field microscope (XLFM), also known as Fourier light field microscope, is a straightforward, single snapshot solution to achieve this. The XLFM acquires spatial-angular information in a single camera exposure. In a subsequent step, a 3D volume can be algorithmically reconstructed, making it exceptionally well-suited for real-time 3D acquisition and potential analysis. 
Unfortunately, traditional reconstruction methods (like deconvolution) require lengthy processing times (0.0220 Hz), hampering the speed advantages of the XLFM. 
Neural network architectures can overcome the speed constraints at the expense of lacking certainty metrics, which renders them untrustworthy for the biomedical realm.
This work proposes a novel architecture to perform fast 3D reconstructions of live immobilized zebrafish neural activity based on a conditional normalizing flow. It reconstructs volumes at 8 Hz spanning 512x512x96 voxels, and it can be trained in under two hours due to the small dataset requirements (10 image-volume pairs). Furthermore, normalizing flows allow for exact Likelihood computation, enabling distribution monitoring, followed by out-of-distribution detection and retraining of the system when a novel sample is detected. We evaluate the proposed method on a cross-validation approach involving multiple in-distribution samples (genetically identical zebrafish) and various out-of-distribution ones.
\end{abstract*}




\section{Introduction}\label{sec:Introduction}
Analysis of fast biological processes on live specimens is a crucial step in biomedical research, where fluorescence 3D microscopy plays an essential role due to its ability to visualize specific structures and processes, either through intrinsic contrast or an ever-grown collection of labeling techniques, including \eg genetically encoded indicators of neuronal activity.

The \gls{xlfm} \cite{Wen2012XLFM}, or \gls{flfmic} \cite{Hua21FLFM, Uo2019FLFM, Han2022FLFM, Stefanoiu2020FLFM}, offers scan-less captures on transparent samples, affordability, and simplicity over scanning microscopes like spinning disk confocal and light sheet (capturing at 10Hz \cite{Hillman2019-otherMicro}). But at the expense of requiring a 3D reconstruction in a post-acquisition step.

Traditionally, reconstructions are done using iterative methods, like the Richardson-Lucy deconvolution \cite{lucy1974iterative}, where the reconstruction quality is sufficient to discern neural activity. However, large computation hardware allocation and long waiting times are required (1 second per iteration in an iterative algorithm), making it impractical for real datasets where thousands of images must be processed.

As an alternative, deep learning approaches arose, where fast reconstructions are possible(up to 50Hz
). These networks are trained on pairs of either raw XLFM or LFM \cite{Levoy2006LFM} images and 3D volumes. Like the XLFMNet \cite{slnet_page2021}, the VCD \cite{Wang2021V2C}, the LFMNet \cite{Page2021LFMNet}, HyLFM-Net \cite{Wagner2021}, and others.

However, within these methods, only the HyLFM-Net can detect network deviations or incapability of handling new sample types, but at the expense of a complex imaging system for continuous validation. An algorithm's lack of certainty metrics renders it unsafe to be integrated in established experimental imaging workflows, as artifacts might be introduced by the network and cannot be detected.

These motivations bring our attention to more statistically informing methods, such as \gls{nfs} \cite{dinh2014niceNF, Danilo2015NF}, a type of invertible neural network recently used for biomedical imaging \cite{Denker2021NF_MI}, inverse problems \cite{ardizzone2018NF_IP, Ardizzone2019NF_IP, Padmanabha2021NF_IP, Shayan2022NF_IP} and, other applications like image generation \cite{WF_jason2020, ardizzone2019NF_IG}. \gls{nfs} learn a mapping between an arbitrary statistical distribution and a normal distribution through a set of invertible and differentiable functions. Also, a tractable exact likelihood allows for probing the quality of this mapping for individual samples, which, in turn, allows for deciding what to do with the new sample, perhaps retraining the network with the new data, if desired.

One disadvantage of \gls{nfs} is that due to the required invertible mapping, no data bottlenecks are possible (like in an encoder-decoder approach) as this would lead to information loss, making it necessary to store all the tensors and gradients in memory during training. This limits the data size that can be used due to the limited \gls{gpu} memory. The conventional solution to this issue is to split the processed tensor after each invertible function \cite{dinh2016RNVP}, feed only one part to the following function, and concatenate the other part to the output tensor of the \gls{nf}. However, when working with large tensors, the gradients during training still overwhelm conventional \gls{gpu}s like the ones found on image acquisition workstations. 

Hence, \gls{wf} \cite{WF_jason2020} was introduced, where an invertible down-sampling operation is used (such as the Haar transform \cite{haar1910wavelets}) to serially down-sample the input image to the desired size (could be down to a single pixel), where a \gls{nf} is used to learn the Haar detail coefficients required to perform up-sampling when performing reconstruction. The last down-sampling step comprises an \gls{nf} that directly learns the probability distribution of the lowest-resolution image.
This allows independent training of each down-sampling \gls{nf} with commercial \gls{gpu}s, allowing their usage for high data throughput for the first time. 
Generating a new image involves running the \gls{wf} backward. The lowest resolution image is sampled from an NF, then up-sampled through all the \gls{wf}s until reaching the original resolution.

The original \gls{wf} approach is not designed for inverse problems as it ignores the image formation model prior knowledge, such as the raw XLFM image or the point spread function.
Furthermore, when training each \gls{nfs}, the Haar transform operator generates a down-sample of the \gls{gt} volume down to the lowest resolution. And when performing a 3D reconstruction, the low-resolution volume received by the \gls{nfs} might slightly deviate from the \gls{gt}. 
These deviations accumulate as the volumes are propagated upwards through the network, affecting the output 3D volume heavily. The quality of the lowest-resolution volume dramatically impacts the final reconstruction, making the lowest-resolution initial guess fundamental to the quality of the full-resolution reconstruction.

\begin{figure}
    \centering
    \includegraphics[width=\textwidth]{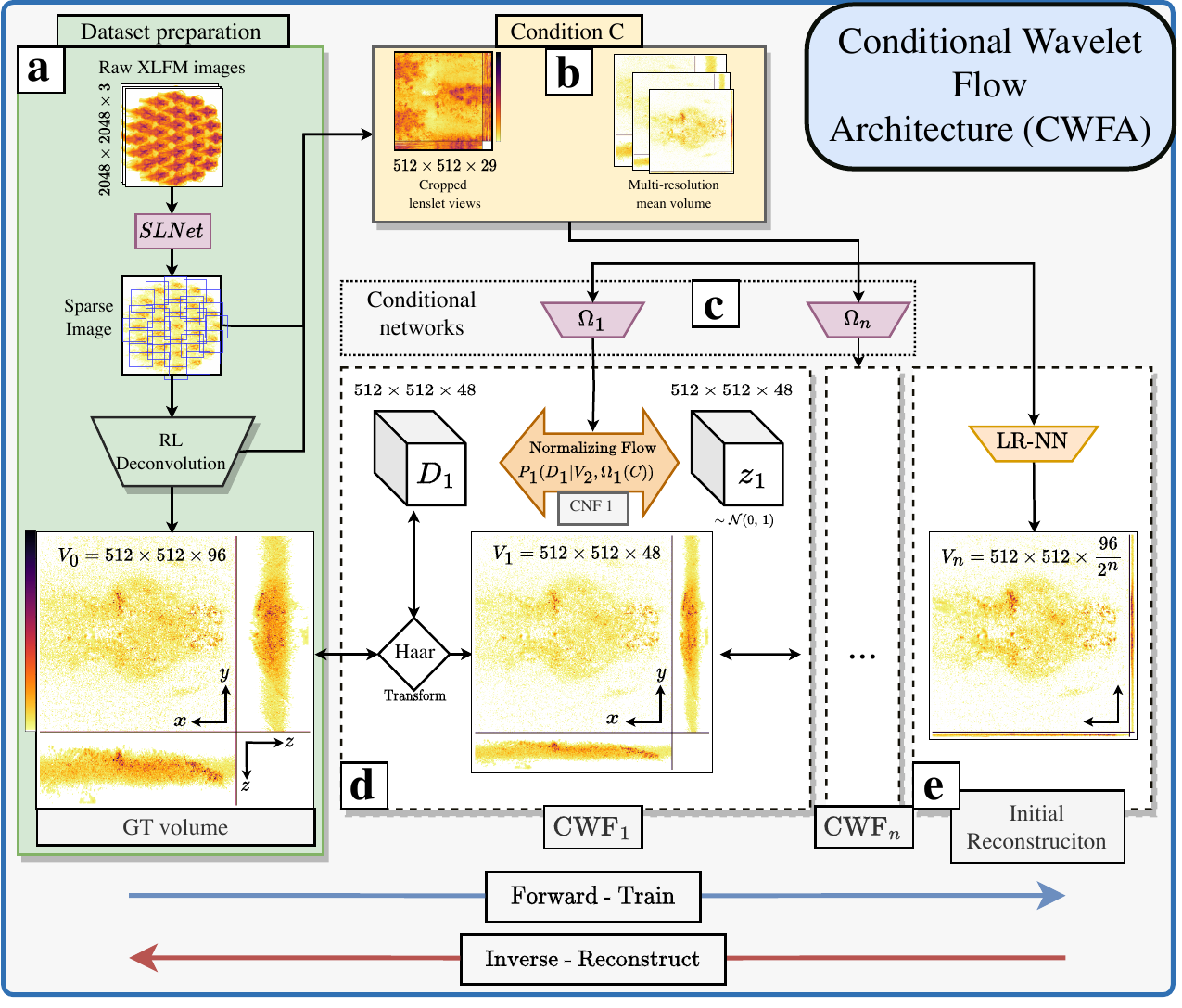}
    \caption{
    The Conditional Wavelet Flow Architecture and workflow: (a) Data preparation, extracting the sparse spatiotemporal signal from the raw XLFM acquisitions with the SLNet\cite{slnet_page2021} and performing 3D reconstructions using the RL algorithm. In (b), the conditions are prepared by cropping and stacking the images and computing the mean of the training volumes. In (d), the full-resolution GT volume $V_0$ and conditions (b) are used to train the \gls{cwf}s. Training is performed in each \gls{cwf} individually and consists in feeding $V_0$ and the processed condition $\Omega_1(C)$ to the \gls{cwf}1. This generates as outputs $V_1$ and $z_1$. The latter is used in loss function \ref{eq:NLLloss}. $V_1$ is fed to the next \gls{cwf}, which is trained similarly. This is repeated until reaching the lowest resolution output. The $\text{LR-NN}$ is trained deterministically with $V_n$ and $C$ pairs.
    }
    \label{fig:CWF}
\end{figure}

Our work proposes the \gls{cwfa}, shown in Fig.~\ref{fig:CWF}, suited for the 3D reconstruction of live immobilized fluorescent samples imaged through an XLFM and \gls{oodd}. It uses \gls{wf}s conditioned by the XLFM-measured image and a 3D volume prior (\eg the mean of the training volumes). And could be easily modified to work on freely behaving animals by removing the dependency on a 3D volume prior.
The \gls{cwf}s reconstruction, and \gls{ood} capabilities enable a fast, accurate, and robust method for 3D fluorescence microscopy. Fast enough to be applied within close-loop or human-in-the-loop experiments, where on the fly analysis of the activity could trigger further steps in an experiment.

In \gls{oodd}, having access to an exact likelihood computation allows for evaluating how likely a sample is to belong to the training distribution at different image scales. Measured in all the down-sampling \gls{cnf}s in the network.

Specifically, in a testing step, the likelihood of a novel sample can be computed by processing it with the \gls{cwf}, as described in sec.~\ref{sec:sub:DSD} and Fig.~\ref{fig:DSA}.

Even though the literature often mentions that \gls{nf}s are not reliable for detecting \gls{ood} samples \cite{kirichenko2020normalizing}, we find that the proposed \gls{cwfa} and sample type enables this capability. 

We demonstrate the reliability of the proposed \gls{cwfa} on XLFM acquisitions of live immobilized zebrafish larvae full-brain neural activity, processed with the SLNet \cite{slnet_page2021}, which separates the neural activity from the background of the acquisitions. 
As a gold standard or \gls{gt} we used 3D reconstructions (100 iterations with the RL algorithm) of the SLNet output.


In sec.~\ref{sec:results} and Fig.~\ref{fig:recon_results}, we compare the \gls{rl} 3D reconstruction (ground truth or golden standard) against the proposed \gls{cwfa}, the XLFMNet \cite{dl:pageXLFMNet2021} and the WF \cite{WF_jason2020} on volumes spanning a FOV of $734 \times 734 \times 225 \mu m^3$ ($512 \times 512 \times 96$ voxels), the proposed \gls{cwfa} operated $735\times$ faster than the traditional deconvolution (1 second per \gls{rl} iteration, we used 100 iterations in this work), with similar quality, as indicated by a mean PSNR of $51.48$, a \gls{mape} \cite{DEMYTTENAEREmape2016} of $0.1011$ or $10.11\%$. 
Alternative methods such as the XLFMNet and \gls{wf} achieved a speed increase of $3200\times$ and $961\times$, but with low-quality reconstructions (PSNR of $47.35$ and $44.30$ and \gls{mape} of $0.3594$ and $0.1167$ respectively.

Furthermore, when analyzing the neural activity of individual neurons in a time sequence, as seen in Fig.~\ref{fig:neural_activity}, the proposed method achieves a \gls{pcc} \cite{benesty2009pearson} of $0.9077$, where the XLFMNet and WF achieve $0.6897$ and $0.8428$ respectively.

In domain shift detection, the first \gls{cwf} performed best and achieved a classification F1-score \cite{f1score92} of $0.988$ and an area under the curve (AUC) of $0.997$, as seen in Fig.~\ref{fig:OOD_example}.
Once an out-of-distribution sample type is detected, we found that 5 minutes of fine-tuning of the proposed \gls{cwfa} on a novel sample are enough to increase quality substantially. Achieving a mean increase of $36.56\%$ in PSNR $40.71\%$ in \gls{mape} and $2.15\%$ \gls{pcc}.

To conclude, the \gls{cwfa} is $735\times$ faster than the reconstruction gold standard and 9\% more accurate than the XLFMNet, a \gls{cnn} previously used for this particular task. The \gls{cwfa} offers excellent domain shift detection that can trigger either re-training or fine-tuning on new data.

\section{Materials and Methods}\label{sec:methods}
\subsection{Tradicional Conditional normalizing flows}\label{sec:sub:CNF}

A \gls{nf}, through a sequence of invertible and differentiable functions, transforms an arbitrary distribution $p_{\boldsymbol X}(\boldsymbol x)$ into the desired distribution $p_{\boldsymbol Z}(\boldsymbol z)$ (usually a normal distribution, hence the name Normalizing Flows). This is possible through the change of variables formula from probability theory, where the density function of the random variable $\boldsymbol X$ is given by:

\begin{equation}
p_{\boldsymbol X}(\boldsymbol x) = p_{\boldsymbol Z}(\boldsymbol z) \lvert \det (J) \rvert,
\label{eq:change_variables}
\end{equation}

Where $\boldsymbol z$ is normally distributed (mean 0 and variance 1), with probability density function $p_{\boldsymbol Z}(\boldsymbol z)$. An NF can be trained by setting $\boldsymbol z = f_\Theta(\boldsymbol x)$, where $f_\Theta$ is an invertible differentiable function parameterized by $\Theta$, and $J = J_\Theta = \frac{\partial f_\Theta(\boldsymbol x)}{\partial \boldsymbol x}$ is the Jacobian of $f_\Theta$ with respect to $\boldsymbol x$, also known as the volume correction term.
As seen in Eq.~\ref{eq:change_variables}, a tractable and easily computable Jacobian determinant of $f_\Theta(\boldsymbol x)$ is preferred (for example, where the Jacobian is block-triangular or diagonal). Hence, choosing the functions conforming $f_\Theta(\boldsymbol x)$ is a crucial and well-studied step \cite{Park_2019_CAT,kingma2018glow,dinh2016RNVP}, out of the scope of this work.

A \gls{nf} can be modified into a \gls{cnf} and represent a conditional distribution $p_{\boldsymbol X}(\boldsymbol X \lvert \boldsymbol C)$, for a set of observations $\boldsymbol X={\boldsymbol x^{(1)},\boldsymbol x^{(2)}, \ldots, \boldsymbol x^{(N)}}$ and conditions $\boldsymbol C={\boldsymbol c^{(1)}, \boldsymbol c^{(2)}, \ldots, \boldsymbol c^{(N)}}$. The likelihood from Eq.~\ref{eq:change_variables} for a single sample $(i)$ becomes:

\begin{equation}
p_{\boldsymbol X}(\boldsymbol x^{(i)} \lvert \boldsymbol c^{(i)},\Theta) = p_{\boldsymbol Z}(\boldsymbol z^{(i)} = f_\Theta(\boldsymbol x^{(i)}, \boldsymbol c^{(i)})) \cdot \lvert \det (J_\Theta^{(i)}) \rvert.
\label{eq:likelihood}
\end{equation}

To train a \gls{cnf}, we need to find the optimal values of the parameters $\Theta$ that maximize the likelihood or minimize the negative log-likelihood of Eq.\ref{eq:likelihood}, defined as follows:

\begin{equation}
\Theta^{*} = \arg \min_{\Theta} \sum_{i=1}^{N} \left[ \frac{||f_\Theta(x^{(i)}, c^{(i)})||^2_2}{2} - \log \lvert \det( J_\Theta^{(i)} ) \rvert + \rho \|\Theta\|^2_2   \right]
\label{eq:NLLloss}
\end{equation}

Where $\det( J_\Theta^{(i)} )$ is the determinant of the Jacobian matrix with respect to $x^{(i)}$, evaluated at $f_\Theta(x^{(i)}, c^{(i)})$. And $\rho \|\Theta\|^2_2$ is the likelihood of the posterior over the model's parameters, assuming a Gaussian distribution weighted by $\rho$.

During training, we minimize the negative log-likelihood function with respect to the parameters $\Theta$ using the Lion optimizer \cite{ref:lion} using weight decay for the parameters posterior. 

After training the \gls{cnf}, we can perform inference on a new image by sampling from the base distribution $p_{\boldsymbol Z}(\boldsymbol z)$ (in our case, a normal distribution $\mathcal{N}(\mu,\sigma^{2})$) and obtaining $\boldsymbol x$ by applying the inverse transformation of the flow: $x=f_\Theta ^{-1} (\boldsymbol z)$. The resulting $\boldsymbol x$ will be a sample from the conditional distribution $p_{\boldsymbol X}(\boldsymbol X \lvert \boldsymbol C)$.
Fig.~\ref{fig:CNF_detail} shows a graphical representation of a \gls{cnf} used for inference.
Even though this method is mathematically sound, its lossless nature limits its usability in practice for large 3D volumes due to high memory requirements.

\subsection{Proposed Conditional Wavelet Flow architecture}
The WF architecture uses a multi-scale hierarchical approach that allows training each up/down-scale independently, allowing flexibility during memory management. Each down-sampled operation is a Haar transform chosen due to its orthonormality and invertibility. Alternatively, to the \gls{wf}s, we used the Haar transform to scale the volumes in the axial dimension, preserving lateral resolution across all steps. The likelihood function of a WF model is given by:

\begin{equation}
    p(V_0) = p(V_n) \prod_{i=1}^{n-1} p(D_i \lvert V_i)
\end{equation}
where $V_0$ is the final high-resolution volume, $D_i$ the Haar transform detail coefficients and $V_i$ the volume down-sampled $i$ times. $p(V_n)$ and all $p(D_i \lvert V_i)$ are normalizing flows.

In the proposed \gls{cwfa}, we substituted $p(V_n)$ with a deterministic \gls{cnn}. We conditioned all the up-sampling \gls{nf}s on a set of external conditions $C$ processed by the \gls{cnn} $\Omega_i$, as shown in Fig.\ref{fig:CWF}. With a likelihood given by:
\begin{equation}
    p(V_0) = p(V_n) \prod_{i=1}^{n-1} p(D_i \lvert \Omega_i(C))
\end{equation}

And the \gls{ll}:
\begin{equation}
    \log p(V_0) = \log p(V_n) + \sum_{i=1}^{n-1} \log p(V_i \lvert \Omega_i(C))
\end{equation}
The final loss function can be optimized independently as $\log p(V_0)$ comprises a sum of probabilities. Showing the advantage of the Wavelet-Flow architecture when using conventional computing hardware.

\subsubsection{Network implementation}
The proposed architecture uses four \gls{cwf}s, each comprised of a Haar transform down-sampling operation and an \gls{cnf}, as seen in Fig.~\ref{fig:CWF}. The internal \gls{cwf} learns the mapping between the Haar coefficients and a normal distribution. And is built from $6$ \gls{cat} blocks (see Supplementary Fig.~\ref{fig:CNF_detail}). 
The number of blocks and their parameters, such as the type of invertible blocks (like GLOW \cite{kingma2018glow}, RNVP \cite{dinh2016RNVP}, HINT \cite{kruse2021hint}, NICE \cite{dinh2014niceNF}, \gls{cat} \cite{Park_2019_CAT} etc.), the number of parameters in each internal convolution, were optimized for \gls{pcc} in a grid-like fashion shown in Fig.~\ref{fig:S:hpoptim}. 
The architecture was implemented in PyTorch aided by the Freia framework for easily invertible architectures \cite{freia}. We invite the reader to explore the source code of this project for further implementation details unde \url{https://github.com/pvjosue/CWFA}. 
The network's total number of parameters is 73 million, where $3.418$ million comprise the \gls{cwf}s, $82K$ the conditional networks $\Omega_{0-n}$ and $63.743$ million the $\text{LR-NN}$. Additional details are in the supplementary section \ref{sec:S:CWFA}. 


\subsection{3D reconstruction: Sampling from the Conditional Wavelet Flow architecture} \label{sec:sub:sampling_CWF}
Fig.~\ref{fig:CWF} depicts how to train the \gls{cwfa} (forward pass) and reconstruct 3D volumes from XLFM images (inverse pass). 
Reconstruction with a trained \gls{cwfa} involves, first, reconstructing a low-resolution volume ($\tilde{V}_n=\text{LR-NN}(C)$), where $\text{LR-NN}$ is a deterministic \gls{cnn} and $C$ the conditions, and using it as input to the \gls{cwf}$_{n-1}$. 
To up-sample on this \gls{cwf}: first, sample $z_{n-1}$ from a normal distribution, and input the pre-processed XLFM image and 3D prior conditions $\Omega_{n-1}(C)$ as a condition to the NF. Then, generate the Haar coefficients ($D_{n-1}$) used for up-sampling $\tilde{V}_{n-1}$ by a factor of 2 on the axial dimension with the Haar transform and generate $\tilde{V}_{n-2}$. 
We repeat this process until reaching $\tilde{V}_0$ at full resolution.
Our architecture comprises $4$ \gls{cwf} and $\text{LR-NN}$. Each \gls{cwf} uses $6$ \gls{cnf}s internally, with $14$ channels per convolution and \gls{cat} invertible blocks, as illustrated in Fig.~\ref{fig:CNF_detail}.
A key parameter during sampling is the temperature parameter, which determines if the $z$ should be sampled from a truncated distribution and to what degree, which is discussed in sec.~\ref{sec:methods:temp}.

\subsection{Input for training} \label{sec:input}
As an input to the network, the \gls{gt} volume at full resolution $V_0$ and the conditions $C$ is required. For both, we pre-processed the raw data with the SLNet, extracting the sparse activity from the image sequences. This approach was chosen due to the fluorescent labeling used (GCaMP), which has only a slight intensity increase ($<10\%$) when intracellular calcium concentration increases as a result of neuronal firing. In other words, the neural activity is practically invisible on the raw sequences that suffer from substantial auto-fluorescence.

\subsection{Conditioning the wavelet flows} \label{sec:conditions}
In the original WF \cite{WF_jason2020}, each NF uses only the next level low-resolution volume as a condition, intending to generate human faces from a learned distribution, where the low-resolution Haar transformed image is used as a condition on each up-sampling step. In such a case, face distribution is the only known information. 

However, when dealing with inverse problems (\eg fluorescence microscopy), prior information about the system and volume to reconstruct are known, such as the forward process in the form of a point spread function (PSF), the captured microscope image and in our case the structural information of a fish, as these are  immobilized with agarose and do not move during the acquisition. 

After the ablation of different configurations (see Fig.~\ref{fig:S:hpoptim}), we chose to use \gls{cat} blocks due to their excellent performance and simplicity. These split the input condition in two, comprised of a translation and a scaling factor (as seen in Fig.~\ref{fig:CNF_detail}) that is applied to the \gls{cwf}'s input in the forward or backward direction. 

The following conditions are fed to each $\text{\gls{cwf}}_i$ after being pre-processed by $\Omega_{i}(C)$:.

\subsubsection{Condition 1: Views cropped from the XLFM image}
This condition acts as the scaling factor of the block and informs the network about neural potential changes.
Due to the nature of the XLFM microscope, each microlens acts as an individual camera, and the 2D image can be interpreted as a multi-view camera problem. 
The raw XLFM input image is prepared first by detecting the center of each micro-lens from the central depth of the measured PSF (using the Python library \textit{findpeaks} \cite{Taskesen_findpeaks_is_for_2020}), then cropping a $512\times512$ area around the 29 centers, and stacking the images in the channel dimension, as seen in Fig~\ref{fig:CWF} panel (b).

\subsubsection{Condition 2: A 3D volume structural prior}
When working with immobilized animals, there's the advantage that only the neural activity changes within consecutive frames. 
Hence, we included this as a prior in the shape of a condition, where we provide to the network a 3D volume created by the mean of the training volumes.
Providing a volumetric prior simplifies the reconstruction problem and allows the network to focus only on updating the neural activity instead of reconstructing a complete 3D volume. Furthermore, as the Haar coefficients along the channel dimension are a discreet derivative of the volume, we found that a processed version of the volume is a very initial approximation, which will be fine-tuned by the \gls{cwf}.

\subsubsection{Conditional networks $\Omega_{0-n}$}
Previous methods using \gls{cnf} used feature extractors such as the first layers of a pre-trained network. 
In this work, we simultaneously trained the conditional networks with the \gls{cwf}s. This, by adding a second data term to the loss function from Eq.~\ref{eq:NLLloss} resulting in the final loss function:

\begin{equation}
    \Theta_i = \Theta^{*}_i + \alpha * \arg \min_{\Theta} \sum_{i=1}^{N} \left[ || V_i - \tilde{V_i})||^2_2\right]
    \label{eq:final_loss}
\end{equation}
Where $\alpha$ is a weighting factor ($0.48$ in this work), $\tilde{V_i}$ is the reconstructed volume at the \gls{cwf} $i$ and $V_i$ is the GT volume down-sampled $i$ times.

In this way, we ensure that the volumes generated by the \gls{cwfa} have not just a statistical constraint but a spatial constraint. 
Without this, the loss might be minimized as the voxel intensities follow the correct distribution, but there is no constraint that, spatially, the reconstructions make sense.

\subsection{Out-of-distribution detection} \label{sec:methods:OOD}
\begin{figure}
    \centering
    \includegraphics[width=\textwidth]{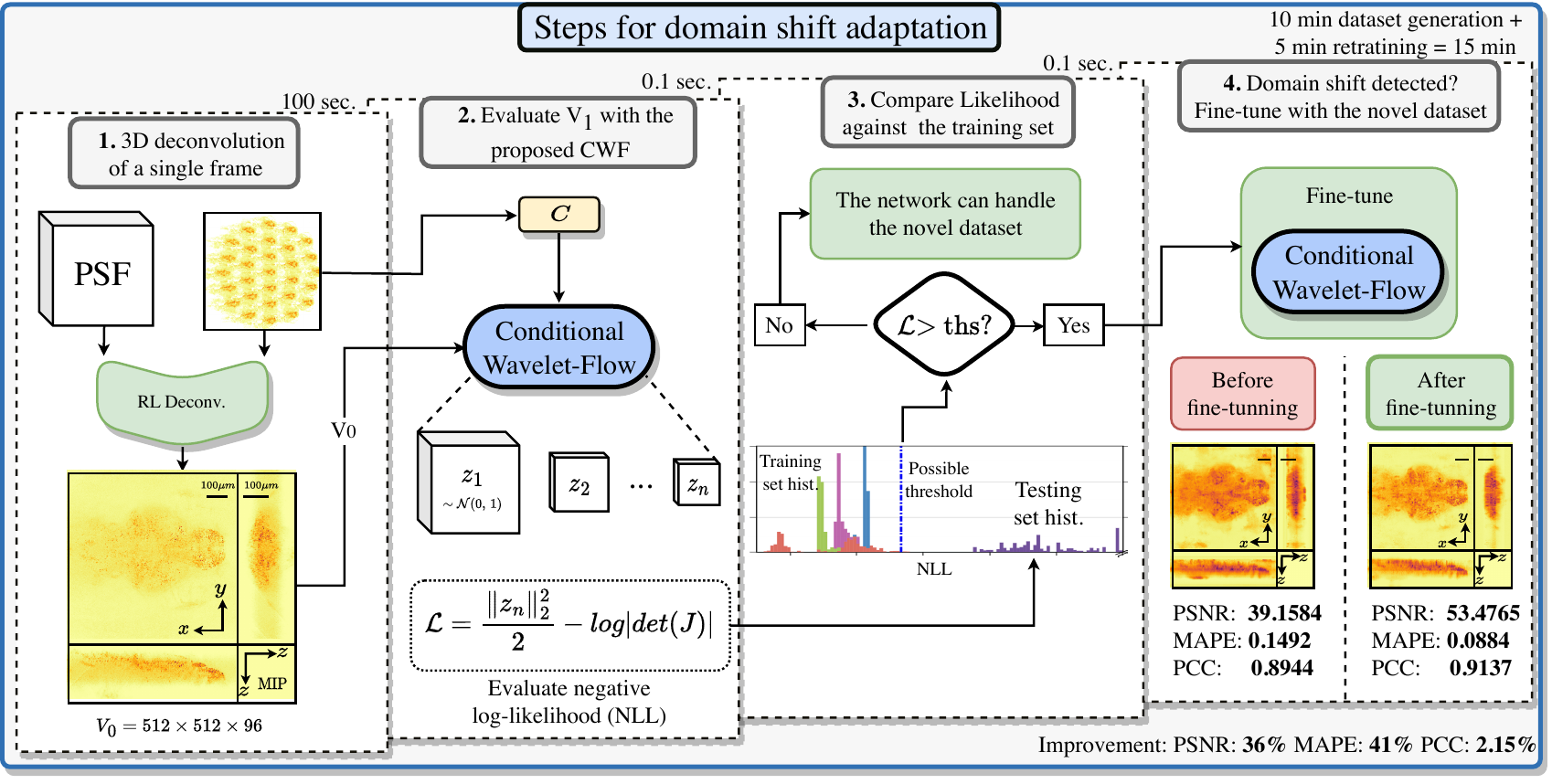}
    \caption{Steps for OOD detection and domain shift adaptation}
    \label{fig:DSA}
\end{figure}

The \gls{oodd} workflow is visually depicted in Fig.~\ref{fig:DSA}. Once we trained a \gls{cwfa} on a set of fish image-volume pairs, we can evaluate if the network can adequately handle a novel sample ($I_{novel}$). 
First, by deconvolving $I_{novel}$ into $V_{novel}$, building the condition $C_{novel}$, and processing these through the network in the forward direction.
Finally, we can evaluate the generated distributions $z_{0-n}$ with eq.~\ref{eq:NLLloss}. 

If the \gls{nll} obtained are above a pre-defined threshold, we have an \gls{ood} sample in hand.
Once detected, there are different solutions, \eg we can fine-tune the \gls{cwfa} on the test sample or add the test sample to the training set. We explore these two options in sec.~\ref{sec:sub:DSadaptation}.

The \gls{oodd} threshold is picked by selecting a small number of samples from all cross-validation training and testing sets and evaluating their likelihood. Then define 1000 thresholds linearly distributed spanning the full range of the data, and pick the one achieving the most significant AUC.

\subsection{Dataset acquisition and pre-processing} \label{sec:sub:dataset}
Pan-neuronal nuclear localized GCaMP6s Tg(HuC:H2B:GCaMP6s) and pan-neuronal soma localized GCaMP7f Tg(HuC:somaGCaMP7f)\cite{ref:shemesh2020precision} zebrafish larvae were imaged at 4–6 days post fertilization. The transgenic larvae were kept at 28°C and paralyzed in standard fish water containing 0.25 mg/ml of pancuronium bromide (Sigma-Aldrich) for 2 min before imaging to reduce motion. The paralyzed larvae were then embedded in agar with 0.5\% agarose (SeaKem GTG) and 1\% low-melting point agarose (Sigma-Aldrich) in Petri dishes. 

Each fish was imaged for 1000 frames at 10Hz. Neural activity images were extracted using the SLNet \cite{dl:pageXLFMNet2021}, as seen in Fig.~\ref{fig:CWF}(a). Later, the resulting images were 3D reconstructed with the RL algorithm for 100 iterations, which takes roughly $1.5$ minutes per frame.

A cross-validation approach was used to evaluate the system, using 6 different fish, each fold trained on 5 different fish and tested on the remaining fish.
For testing \gls{oodd} we used the testing fish on each fold, non-sparse images (deconvolutions of raw data, without pre-processing of the SLNet), and fluorescent beads images.

Ten XLFM pre-processed images and volumes per dataset were used to train the \gls{cwfa}, as each cross-validation set has 5 training datasets; in total, 50 images were used for training, 250 for validation, and 50 for testing.
We tested different amounts of data used for training (5, 10, 20, and 50 pairs), where using 10 images dealt the best performance and a training time of 1:43h, see supplementary Fig.~\ref{fig:S:temp_length} for details.

The beads dataset was created by imaging 1-$\mu m$-diameter   green   fluorescent   beads   (Ther-284moFisher) randomly distributed in 1\% agarose (low melting285point agarose, Sigma-Aldrich). The stock beads were serially diluted  using  melted  agarose  to  $10^{-3}$,  $10^{-4}$,  $10^{-5}$,  $10^{-6}$ of  the original  concentration.

\section{Experiments}\label{sec:results}
\subsection{3D reconstruction of sparse images with the CWFA}

We compared the proposed method against the XLFMNet \cite{slnet_page2021} aiming to match the amount of parameters (103M) and a modified version of the Wavelet Flow \cite{WF_jason2020} (WF). The XLFMNet is a U-net-based architecture \cite{ref:unet} that achieves high reconstruction speeds but lacks any certainty metrics (as in conventional deep learning techniques). 

In the original WF, the lowest resolution was reconstructed from an unconditioned normalizing flow based purely on the training data distribution, which does not comply with inverse problems modus-operandi, where a measurement is required to reconstruct the variable in question. 
The modified version follows the original design in which only the low-resolution image is used as a condition in each flow; however, we used a \gls{cnn} $\Omega_n$ instead of the lowest-resolution NF, informing the system about the measurement and making a fair comparison. We used similar settings optimized for \gls{pcc} as the proposed approach (Fig.~\ref{fig:S:hpoptim}).

\begin{figure}[h]
    \centering
    \includegraphics[width=0.95\textwidth]{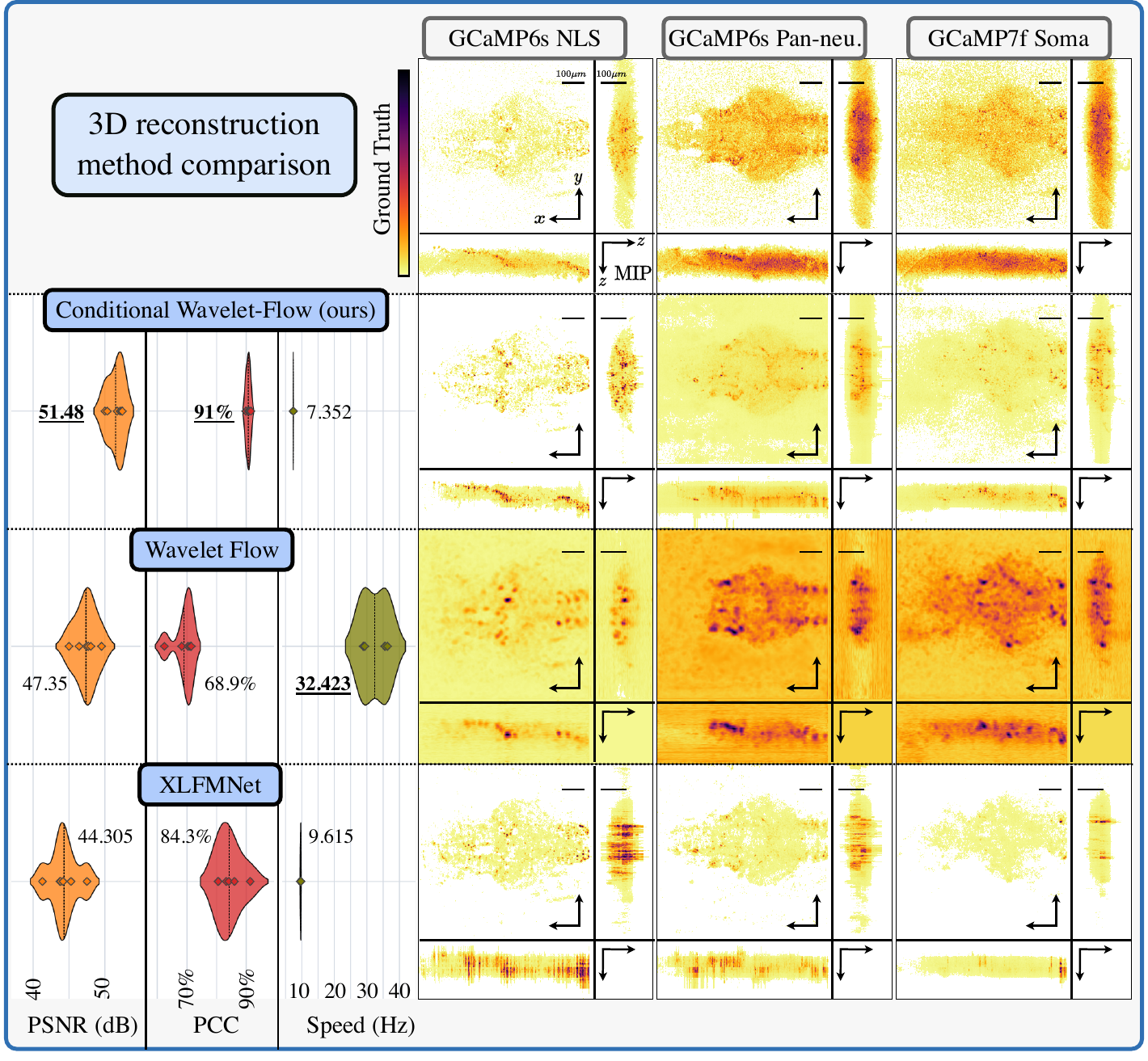}
    \caption{3D reconstruction comparison of zebrafish images with different methods. On the top row, \gls{gt} volumes were generated through 100 iterations of the RL algorithm using a measured PSF. On each column, a different zebrafish sample. The following rows show reconstructions with different methods. The left-most column shows the performance metrics used for comparison: \gls{mape}, followed by the mean \gls{pcc} of 50 frames from the same fish acquisition measured on the top 50 most active neurons per fish. The bottom arrows show the direction of better performant metrics.}
    \label{fig:recon_results}
\end{figure}

\subsubsection{Comparison metrics} \label{sec:sub:metrics}
We identified three relevant aspects for quality evaluation of a 3D reconstruction:
\begin{itemize}
    \item General image quality: We used peak signal-to-noise ratio (\gls{psnr}) as it compares the pixel-wise quality of reconstruction against the \gls{gt}.
    \item Sparse image quality: As the volumetric data used is highly sparse (mostly zeros), we created a mask of the non-zero values on both the \gls{gt} and reconstructions, and, used the mean absolute percentage error (\gls{mape}) for single frame quality assessment
    \item Temporal consistency: An important aspect is that the neural activity is correctly reconstructed across frames. Hence, we used the Pearson correlation coefficient (\gls{pcc}) of single neurons across time. The neuron positions on each data sample were determined with the suite2p framework \cite{suite2p}.
\end{itemize}

As seen in Fig.~\ref{fig:recon_results}, the proposed method outperforms the other two approaches regarding these metrics. However, XLFMNet provides faster inference capabilities but lacks any certainty metrics.

\subsubsection{Sampling temperature} \label{sec:methods:temp}
In our experiments, we found that a temperature of zero produced the highest quality reconstruction. This might be the optimal parameter as the \gls{gt} used is the result of the RL algorithm, which converges towards the maximum-likelihood estimate. A zero temperature means the network reconstructs the most likely sample, which makes sense for an inverse problem. Furthermore, zero temperature means no sampling is required, increasing the network performance. A comparison of different sample temperatures can be found in the supplementary Fig.~\ref{fig:S:temp_length}.

\subsection{Domain shift or out-of-distribution-detection (OODD)} \label{sec:sub:DSD}
We evaluated our method by training the \gls{cwfa} on six cross-validation folds of zebrafish fluorescent activity datasets pre-processed with the SLNet extracting the neural activity. 
Then, presented to the pipeline with different sample types, such as previously unseen fish sparse images (processed with the SLNet), raw XLFM fish images (not pre-processed), and fluorescent bead images with different concentrations. As shown in Fig.~\ref{fig:OOD_example}.

Our algorithm achieves across all cross-validation folds a mean AUC of $0.9964$ and F1-score of $0.9916$ on $\text{\gls{cwf}}_1$, with a \gls{nll} threshold of $-1.33$. The achieved AUC and F1-scores for all the down-sampling \gls{cwf}s are presented in table \ref{table:S:ood_all_steps}.

The \gls{ood} \gls{nll} threshold was established by computing the ROC curve on all cross-validation sets simultaneously and choosing the threshold with the highest F1-score and AUC, in our case $-1.33$.

\subsection{Domain shift adaptation through fine-tuning} \label{sec:sub:DSadaptation}
\begin{figure}[h]
    \centering
    \includegraphics[width=\textwidth]{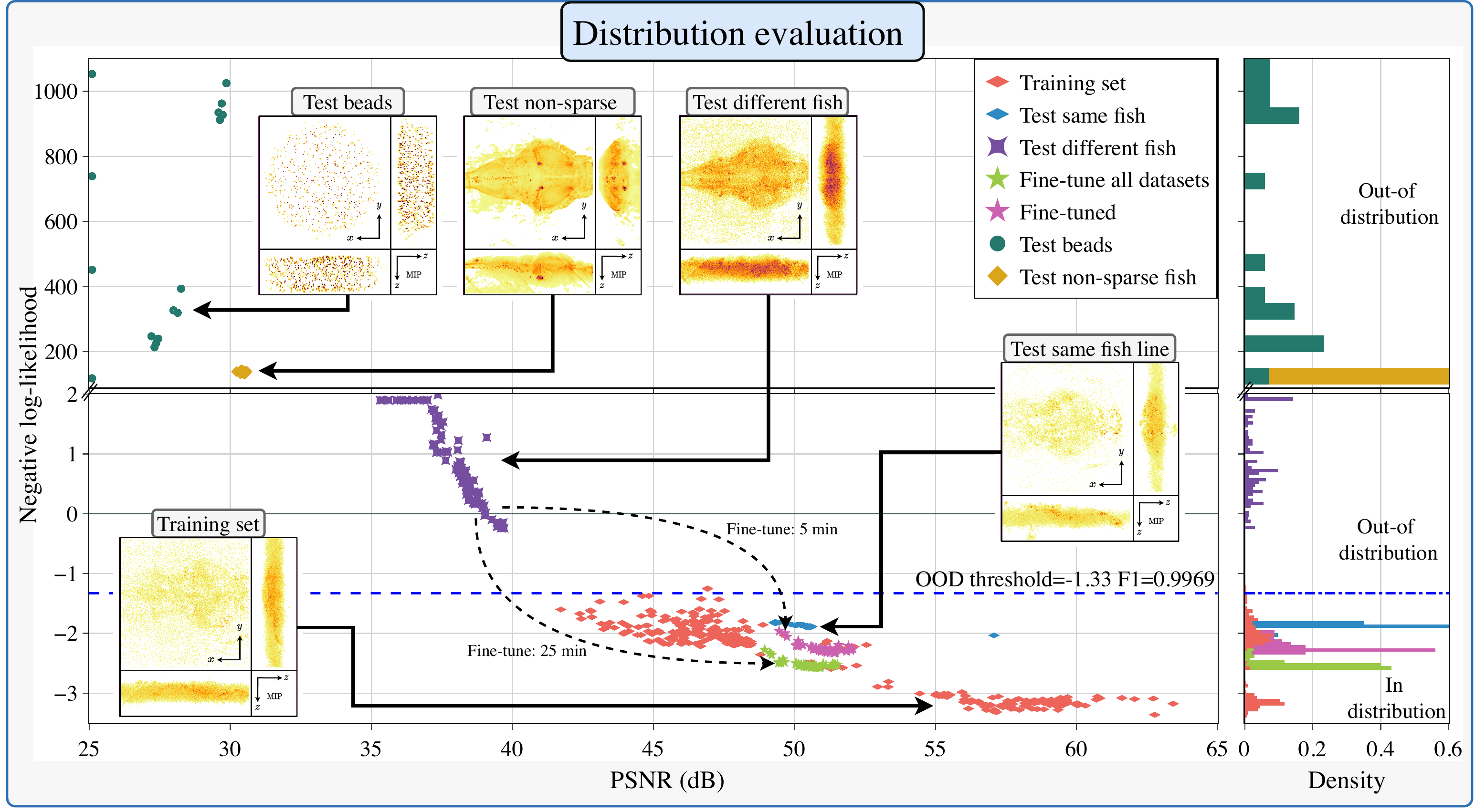}
    \caption{
    Distribution analysis of different sample types. The negative log-likelihood vs. PSNR of the first \gls{cwf} module is on the left panel. Different sample types and a \gls{ood} threshold are presented, used to determine if the network can handle a sample reliably. For the case of the 'Test different fish', we present two re-training approaches, as mentioned in sec.~\ref{sec:sub:DSadaptation}.
    And on the right panel, the \gls{nll} density, where it can be seen that a threshold can be found to separate in vs. out of distribution data.
    }
    \label{fig:OOD_example}
\end{figure}

Once a sample is detected as \gls{ood}, a couple of possible approaches are:
\begin{itemize}
    \item\textbf{Fine-tune the pre-trained network on the new data:} We first need to deconvolve images from the new sample, where each deconvolution takes around 1 minute. Then fine-tune the testing set on each of the cross-validation folds with 10 image-volume pairs for 100 epochs (20 epochs per step). This took roughly 10 minutes to generate the training pairs and 5 minutes for fine-tuning. Dealing a mean increase of $36\%$ on PSRN, $40\%$ on MAPE, and $2\%$ on\gls{pcc}, as seen in table~\ref{tab:sup:perf_increase}, and on Fig.~\ref{fig:OOD_example}, where we fine-tuned the 'Test different fish' into 'Fine-tuned.'
    \item\textbf{Append new data to cross-validation fold:} If we still want to use the network for the fish it already trained on, we can append the new training data to the training set, and fine-tune all the data. This approach takes roughly 10 minutes to generate the training pairs and 25 minutes for fine-tuning. Dealing a mean increase of $34\%$ on PSRN, $34\%$ on MAPE, and $1\%$ on\gls{pcc}, as seen in Fig.~\ref{fig:OOD_example} as 'Fine-tuned all datasets'. 
\end{itemize}

\section{Discussion}
In this work, we presented a Bayesian approach to 3D reconstruction of live immobilized fluorescent zebrafish, comprised of a robust workflow for inverse problems, particularly when false positives should be minimized, as in the case of bio-medical data. 

Remarkably, the amount of data (10 images per sample) and training time (around 2 hours) combined with the \gls{oodd} capability would allow this system to be integrated into a downstream analysis workflow. And when a new fish or fluorescent sample needs analysis, 5 minutes of retraining would suffice to allow the network to reconstruct neural activity reliably.

There are some remaining questions that we leave for future work. Such as, which element in the setup enables the \gls{oodd}. We have some hints, but that would require further experimentation. For instance, the fact that the classification capability increases in higher resolution \gls{cwf} steps (Table.~\ref{table:S:ood_all_steps}) might indicate that the hierarchical approach based on the Haar transform aids the likelihood clustering. It also might be the type of data and initial distribution $P(x)$ (usually Poisson distributed due to fluorescence) is relevant. 

We find that \gls{nf}s and Bayesian approaches are adequate for bio-medical imaging due to their potential capability of handling uncertainty and correctly presenting it to the user, allowing for better decision-making and false positive minimization. 


\section{Backmatter}


\begin{backmatter}
\bmsection{Funding}

J. Page Vizcaíno is supported by the Deutsche Forschungsgemeinschaft (LA 3264/2-1).
Z. Wang, P. Symvoulidis and E. S. Boyden acknowledge the Alan Foundation, HHMI, Lisa Yang, NIH R01DA029639, NIH 1R01MH123977, NIH R01MH122971, NIH RF1NS113287, NSF 1848029, NIH 1R01DA045549, and John Doerr.
P. Favaro acknowledges the interdisciplinary project
funding UniBE ID Grant 2018 of the University of Bern.

\bmsection{Disclosures}
The authors declare no conflicts of interest.





\bmsection{Data Availability Statement}
The sourcecode and dataset of this project can be found under \url{https://github.com/pvjosue/CWFA} and \href{https://doi.org/10.5281/zenodo.8024696}{10.5281/zenodo.8024696} respectibly

\bmsection{Supplemental document}
See Supplement 1 for supporting content. 

\end{backmatter}

\bibliography{Optica-template}

\makeatletter\@input{xx.tex}\makeatother
\end{document}


\maketitle


\section{The Conditional Wavelet flow architecture}
\label{sec:S:CWFA}
We augmented the Freia framework \cite{freia} for better handling of our problem in hand. 
Some of the modifications are:
\begin{itemize}
    \item Permutation along random dimensions: The permutation operation is typically done only on the channel dimension. We modified the implementation to randomly select a dimension to permute and apply a random permutation of the elements on that dimension.
    \item 1D Haar transform: The original Wavelet-Flow paper \cite{WF_jason2020} up/down sampled the images on the x-y axes, however in the case of XLFM images, the lenslet images already contain the high definition information in x-y dimension. Hence, compressing the images to upsample them again seemed counter intuitive. Instead, we perform the up/down sampling along the channel dimension. We are starting with 8 depths and upsampling with 4 CWFs until reaching 96 depths.
\end{itemize}

Additionally, instead of the traditional permutation functions, where the channel dimension is permuted, we included permutations where other dimensions can be permuted. See FIg.~\ref{sec:S:CWFA} for details on the internal components of the network.

\begin{figure}
    \centering
    \includegraphics[width=\textwidth]{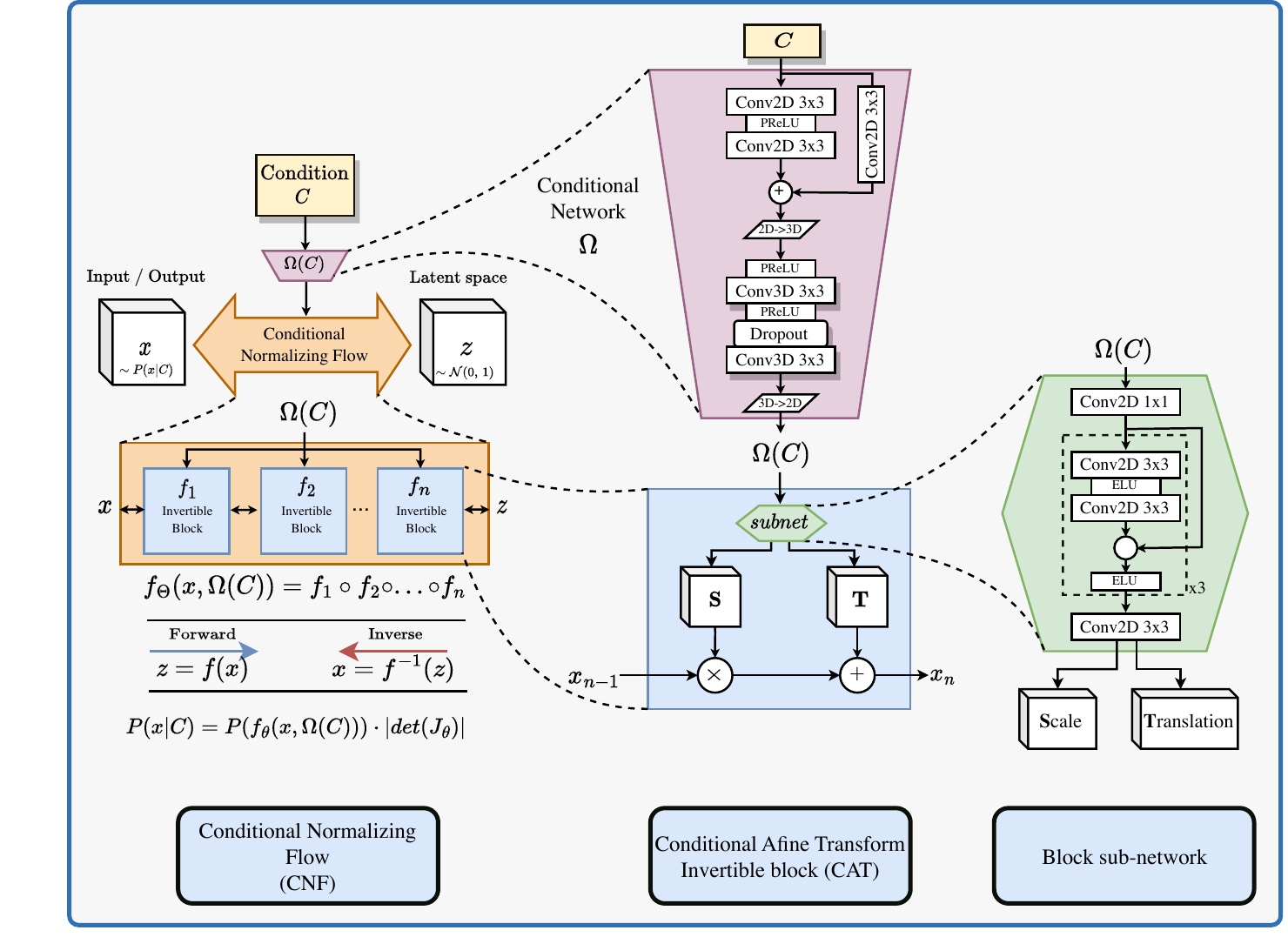}
    \caption{Single conditional normalizing flow used within the CFWA, also present in Fig.~\ref{fig:CWF} as CNF1,2, etc. In blue, the CAT block is responsible for computing a scaling and translation from the condition and applying it to the input}
    \label{fig:CNF_detail}
\end{figure}

\begin{figure}
    \centering
    \includegraphics[width=\textwidth]{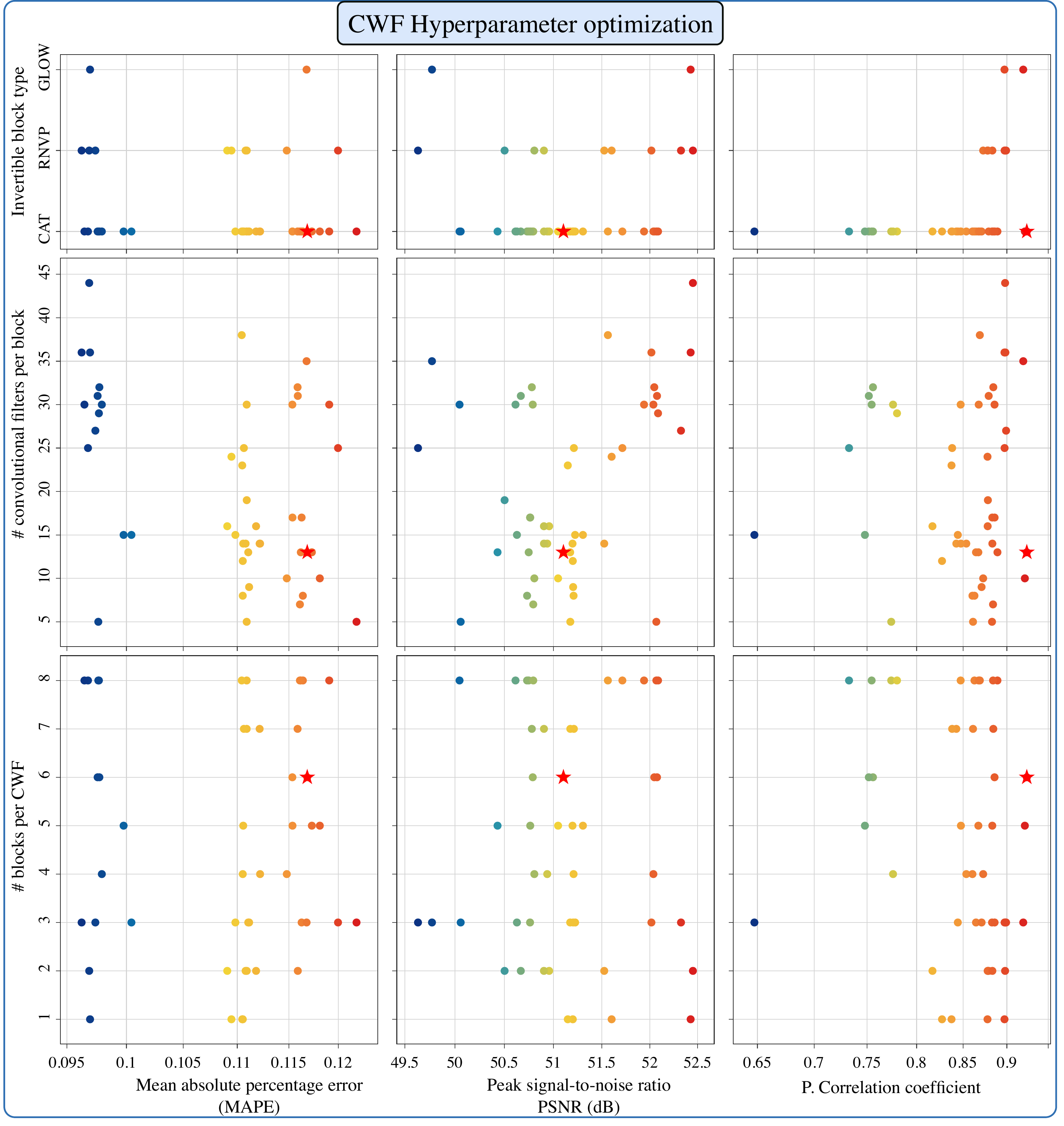}
    \caption{Hyper-parameter ablation optimized towards Pearson correlation coefficient. Highlighted with a red star is the selected architecture.}
    \label{fig:S:hpoptim}
\end{figure}

\section{Low resolution 3D reconstruction (LR-net)} \label{sec:S:LRnet}
$\text{LL-net}$ shown in Fig.~\ref{fig:LR-NN}, performs a low axial resolution reconstruction with an XLFM image and low-resolution mean volume as input. It is comprised of the following parts:
\begin{itemize}
    \item The perspective views corresponding to each microlens is cropped and stacked on the channel dimension.
    \item The views are passed to a 2D unit, where the channel dimension encodes the volume's axial dimension.
    \item In parallel, the mean volume is passed through two ConvNext convolutional blocks and a global attention module in a residual multiplicative matter.
    \item The global attention module selects the relevant pixel-wise information by first turning the image into a 1D array, then applying a Conv1D, a ReLU, a Conv1d, and a sigmoid activation which outputs a value from 0 to 1 that gets multiplied or weights the original pixel. 
    \item the U-net and ConvNext paths are added into a single volume. Resulting in a $512\times512\times \frac{96}{2^n}$ voxels, where $n$ is the number of down-sampling steps in the CWFA (5 in our case).
\end{itemize}

\begin{table}[]
    \centering
    \begin{tabular}{|p{50mm}|c|c|}
        \hline
        \rowcolor{lightgray!100} \textbf{Description} & \textbf{Age} & \textbf{Cross-val fold test}\\
        \hline
        NLS GCaMP6s & unknown age & 0\\
        \hline
        NLS GCaMP6s & unknown age & 1\\ 
        \hline
        Pan-neuronal nuclear localized GCaMP6s Tg(HuC:H2B:GCaMP6s) & 4-6 days & 2\\ 
        \hline
        Pan-neuronal nuclear localized GCaMP6s Tg(HuC:H2B:GCaMP6s) & 4-6 days & 3\\ 
        \hline
        Soma localized GCaMP7f Tg(HuC:somaGCaMP7f) & 4-6 days & 4\\ 
        \hline
        Soma localized GCaMP7f Tg(HuC:somaGCaMP7f) & 4-6 days & 5\\ 
        \hline
    \end{tabular}
    \caption{Description of fish used for this work and in which cross-validation set they were used as testing set.}
    \label{table:used_fish}
\end{table}

\begin{table}[h]
    \centering
    \setlength{\tabcolsep}{3pt}
    \renewcommand{\arraystretch}{1}
    \begin{tabular}{|c|c|c|c|c|c|}
       \hline
         \rowcolor{lightgray!100} \textbf{Metric} &  \textbf{Before fine-tune} & \textbf{Fine-tune only testing DS} & \textbf{$\% \Uparrow$} & \textbf{Fine-tune all DSs} & \textbf{$\% \Uparrow$} \\
         \hline
         \hline
         PSNR   & 39.1584   & 53.4765   & 36.5644\% & 52.6036 & 34.3352\% \\
         MAPE   & 0.1492    & 0.0884    & 40.7160\% & 0.0984  & 34.0131\% \\
         PCC    & 0.8944    & 0.9137    & 2.1578\%  & 0.9065  & 1.3528 \\
         Time   & -         & 5 min.        & -      & 25 min.      & - \\
         \hline
    \end{tabular}
    \caption{Improvement upon fine-tuning on 10 images of the testing set (column 3-4). Or by appending the 10 images to the cross-validation training set (column 5-6).}
    \label{tab:sup:perf_increase}
\end{table}


\begin{figure}
    \centering
    \includegraphics[width=\textwidth]{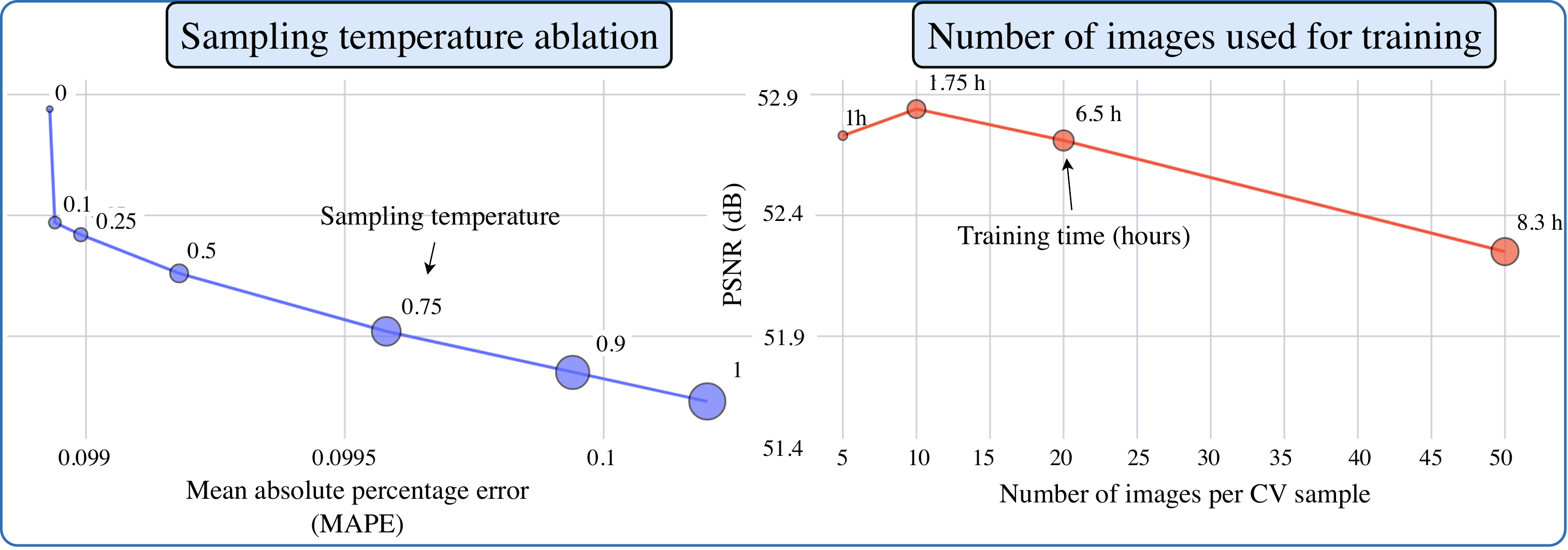}
    \caption{(Left) Sampling temperature ablation. (Right) The number of samples used per CV set during training. Both were evaluated on CV fold 0. We performed a similar evaluation fo the WF parameters and the LR-Net.}
    \label{fig:S:temp_length}
\end{figure}

\begin{figure}
     \centering
     \begin{subfigure}[b]{0.45\textwidth}
     \centering
            \includegraphics[width=\textwidth]{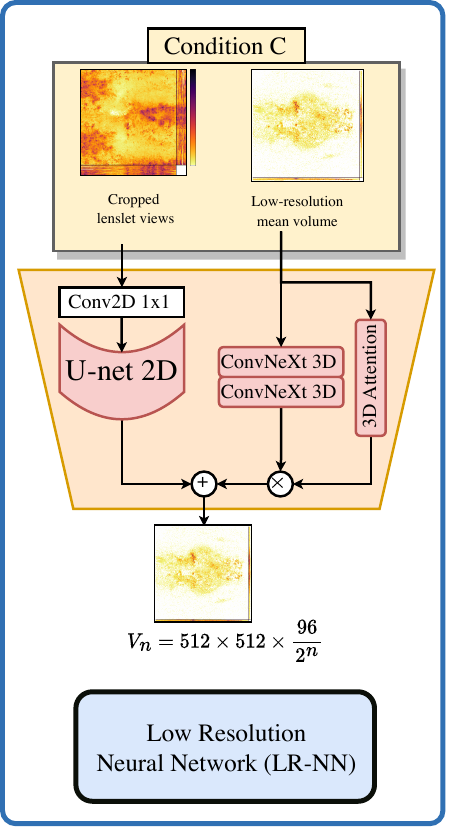}
            \caption{Low-resolution reconstruction network, clarified in sec.~\ref{sec:S:LRnet}, used as an initial step for the reconstruction operation. Where the stacked lenslet views of the 2D XLFM image and a mean low-resolution volume are used as input. And a 3D volume with low axial resolution results as an output. Later upsampled by the remaining CWF blocks in the CWFA network.}
            \label{fig:LR-NN}
     \end{subfigure}
    \hfill
     \begin{subfigure}[b]{0.45\textwidth}
            \centering
            \setlength{\tabcolsep}{3pt}
            \renewcommand{\arraystretch}{1}
            \begin{tabular}{|c|c|c|c|}
               \hline
                 \rowcolor{lightgray!100} \textbf{CWFA step} &  \textbf{NLL threshold} & \textbf{F1-score} & \textbf{AUC} \\
                 \hline
                 \hline
                 1 & -1.33 & 0.9933 & 0.9985\\
                 2 & -0.65 & 0.9886 & 0.9933\\
                 3 & -0.21 & 0.9402 & 0.9647\\
                 4 & -0.26 & 0.9615 & 0.9833\\
                 \hline
            \end{tabular}
            \caption{Out-of-distribution detection scores for all CWF steps.}
            \label{table:S:ood_all_steps}
     \end{subfigure}

\end{figure}

\begin{figure}[h]
    \centering
    \includegraphics[width=\textwidth]{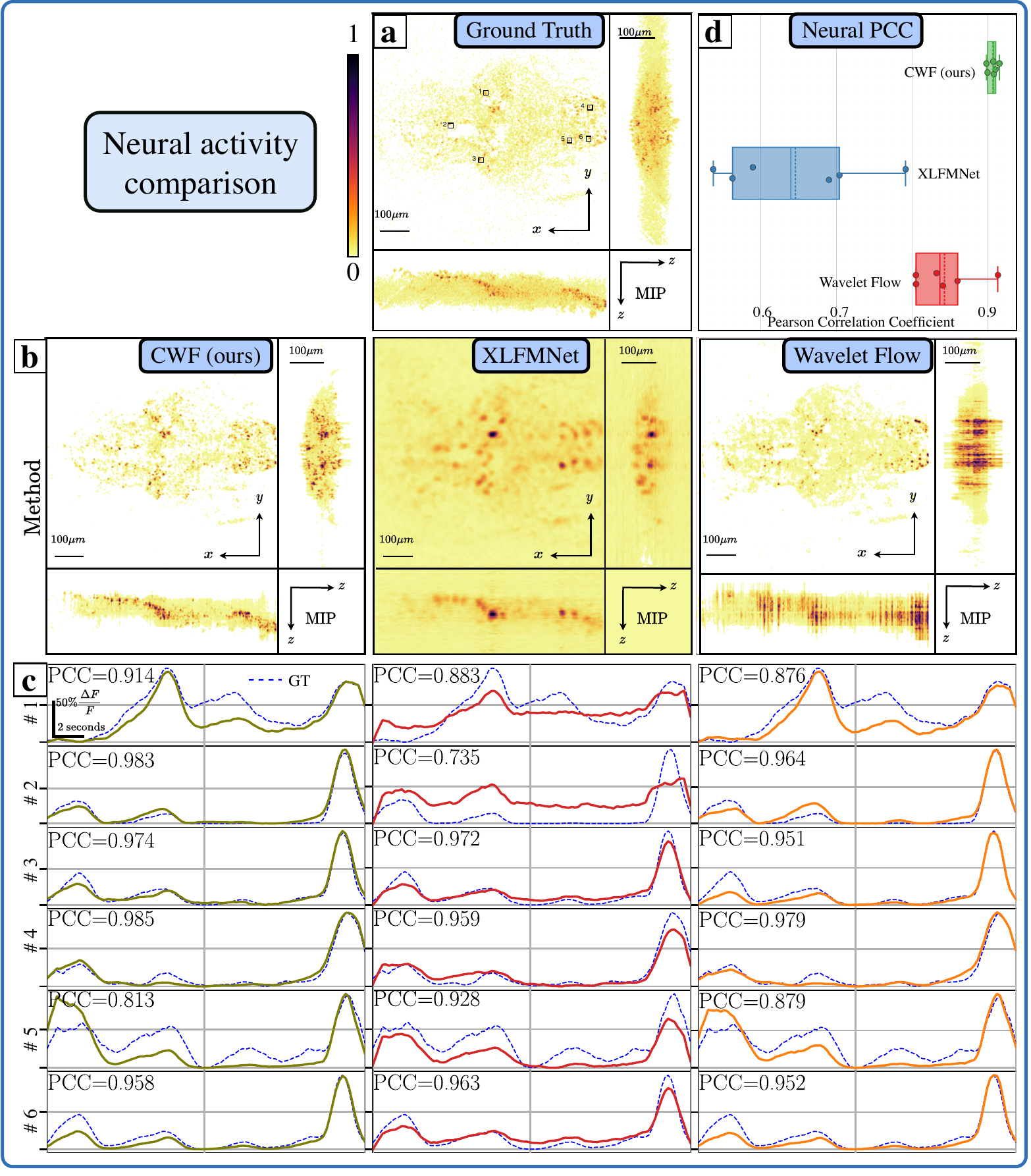}
    \caption{Neural activity comparison with different methods. In \textbf{(a)}, the MIP of the GT volume, with a subset of the active neurons highlighted. Followed by a reconstructed frame with different methods in row \textbf{(b)}. In \textbf{(c)}, the neural potentials of 6 neurons in 100 frames (10 seconds). (d) the mean Pearson correlation coefficient with six fish and the three methods are shown. Note that the PCC was measured directly on the 3D volumes, and only the 2D projection of the neuron coordinates are shown in \textbf{(a)}}
    \label{fig:neural_activity}
\end{figure}

